# Is there a chance that the anisotropies in the cosmic microwave background detected by COBE are an artefact of COBE's image reconstruction?

Keith S Cover[1]
VU University Medical Center, Amsterdam[1]

Corresponding author:
Keith S Cover, PhD
VU University Medical Center
Post Box 7057
1007 MB Amsterdam
The Netherlands
Email: Keith@kscover.ca
*Tel:*   31 20 444-0677
Fax:   31 20 444-4816

**Abstract**

The COBE time ordered data (TOD) was reanalysed for anisotropies in the cosmic microwave background with a novel technique that is much simpler and more robust than used in the official analysis of the COBE TOD. The technique extends the statistical concept of rejecting the null hypothesis to image reconstruction by asking if the TOD is consistent with no anisotropies outside the galactic band. If it is assumed that the instrumentation noise is ideal, as appears to be the case in the official analysis, the null hypothesis can be rejected to $28\sigma$, a highly significant detection. However, the official analysis presents no information supporting the ideal noise assumption to the level required to reject the null hypothesis. The interaction of COBE's calibration error of roughly 3.0% with the brightest regions of the galactic band in combination with the reconstruction algorithm may have introduced artefacts roughly of the size and spectral signature of the 10ppm anisotropies reported by COBE. Data from follow up missions have similar problems with calibration errors that may have resulted in similar sky maps to COBE. It is pointed out that either a few hours of repetitive circular scans using the WMAP satellite or the single beam antenna of the soon-to-be-launched Planck satellite should provide definitive evidence as to the existence or non existence of the 10ppm anisotropies without the use of image reconstruction.

**Keywords** techniques: image processing, (cosmology:) cosmic microwave background

# 1 INTRODUCTION

Precise and accurate measurement of anisotropies in the cosmic microwave background (CMB) would provide a treasure trove of information on the cosmos. As a consequence, tremendous effort has been expended on trying to detect and measure any anisotropies. These efforts have included two satellite missions that have published results, the Cosmic Background Explorer (COBE) (Smoot et al. 1992; Bennett et al. 1996A) and the Wilkinson Microwave Anisotropy Probe (WMAP) (Bennett et al. 2003; Spergel et al. 2003). In addition, the Planck satellite is scheduled to be launched in 2008 (Tauber 2004).

The official COBE analysis reported detecting anisotropies at roughly 10 ppm of the CMB temperature. The WMAP mission closely followed the COBE acquisition and reconstruction strategy but improved on the spatial resolution and reported detecting similar anisotropies. Both COBE and WMAP used a differential microwave radiometer (DMR). Rather than directly measuring the intensity of the CMB at each point in the sky, the DMR's measured the difference between two widely separated beams. For COBE the separation was 60˚. For each satellite, a sky map of the fluctuations in the intensity of the CMB at each point in the sky was then reconstructed by finding the sky map that was most consistent with the DMR measurements in the maximum likelihood sense.

A major impediment to measuring the anisotropies in the CMB is the dust concentrated near the galactic plane. As the dust at the galactic centre is up to 1000 times brighter than the reported anisotropies in the CMB (Barnes et al. 2003), great care was taken during the official COBE analysis to ensure the dust signal did not contaminate the measurements outside the galactic band. However, the complexities and uncertainties of COBE's instrumentation make this a daunting task. While most of the dust lies within 20° of the galactic plane, at some longitudes it extends several degrees further (Kogut et al. 1996B).

An inherent problem in designing any reconstruction algorithm is dealing with the nonuniqueness of the result – the fact that many different sky maps can be consistent with the same data set (Press et al. 1992). The official COBE analysis dealt with the nonuniqueness by using the sky map with the maximum likelihood fit to the data. Therefore, of all the sky maps consistent with the data, the sky map that fit the data with the smallest $\chi^2$ measure (Press et al. 1992) was used. Thus, the official analysis overlooked the variety of other sky maps that were also consistent with the COBE TOD set to within the confidence level of $\chi^2$, and the additional uncertainty this adds to the reliable detection of the anisotropies.

By all accounts the COBE team put tremendous time, care and effort into ensuring the reported detection was as reliable as possible (Kolgut et al. 1992, 1996A). However, the author can find nothing in the literature to suggest that anyone applied statistical tests for the existence of the anisotropies that took into account the full range of sky maps consistent with the TOD. While it is a common practice when interpreting reconstructed images to not explicitly take the reconstructed nature of the images into account, and instead use statistical measures intended for direct measurements, it leaves the analysis incomplete and prone to confusing and misleading results.

A possible source of false positive signals that can be compounded by image reconstruction is non ideal noise. Ideal noise has tidy statistical properties including the well known rule that the standard deviation of the noise reduces by the square root of the number of averages. COBE used averaging extensively to improve the signal to noise (SNR) of the data by more than a factor of a thousand. The roughly 3.0% uncertainty in the calibration (also known as gain drift) of the COBE TOD over the 4 year mission may have caused a significant violation of the ideal noise assumption as the signal from the galactic band and the instrumentation noise were modulated by the error in the calibration (Kolgut et al. 1996A). (The uncorrected calibration error was near 3%. The calibration error was corrected during the official COBE analysis and the corrected calibration error was estimated to be 0.7%.) Also, while the maximum likelihood reconstruction algorithm is normally linear, the official COBE analysis modified it in a nonlinear manner than may have been particularly sensitive to the non ideal noise. In addition, estimation of the error in the official analysis of the COBE used simulations that seemed to have assumed ideal noise.

In the official COBE analysis the gain drift of the TOD was not considered a significant problem as long as it was slow with respect to the 75s rotation rate of the satellite (Kogut et al, 1992). However, the interaction between the nonlinear gain drift and the linear reconstruction algorithm can be quite complicated making its effects difficult to predict. Importantly, the reconstruction algorithm did not take into account the timing of the data acquisition. The reconstruction largely assumed that all observations over the 4 years of acquisition were acquired at the same point in time.

The selection of the best statistical test for the detection of a signal in a reconstructed image is an unresolved issue in many fields. For example, the multiexponential reconstruction problem is much simpler to state mathematically than COBE's but is also nonunique. There is no consensus on how to reconstruct multiexponential decays. A particular problem in multiexponential reconstruction, the myelin water peak detection problem, is surprisingly similar to the CMB anisotropy problem. Both techniques are trying to detect a small signal (the anisotropies) which is adjacent to a much larger one (the galactic band) and both problems are approached using reconstruction algorithms. Indeed, the idea of generalising the concept of rejecting the null hypothesis to image reconstruction and the modified maximum likelihood algorithm used in this paper were originally designed by the author for the multiexponential reconstruction problem of in vivo myelin water detection (Cover 2006).

For the COBE reconstruction problem, to test the null hypothesis that the anisotropies outside the galactic band were zero, a reanalysis was completed using a modified maximum likelihood reconstruction algorithm that forced all pixels to zero that were beyond a specified number of degrees from the galactic equator. The specified number of degrees was then varied from 90° to 0° and the best fit by the $\chi^2$ measure was calculated for each. The sky maps and $\chi^2$ values were then examined to determine if the null hypothesis could be rejected to any significant confidence level.

This paper is not the first to raise questions about the results reported by COBE and WMAP. Concerns about the unusual characteristics of some of the results have been raised previously (Starkman and Schwarz 2005). Nevertheless, it is essential for the progress of science that the fundamental basis for assessing the reliability of any experimental results be the quality of the experimental technique and analysis rather than the consistency of the results with various theories.

## 2. THEORY

When trying to detect a signal with direct measurement - such as by the soon-to-be-launched Planck satellite - the question is whether the measured value is sufficiently different from zero. The statistical test used is generally presented as a "null hypothesis" that must be rejected to a sufficient confidence level. In the Planck direct measurement case, the null hypothesis is that the measured value is zero. Assuming the measured value has ideal noise, the confidence level is often specified in terms of the number of standard deviations of the noise, with higher values indicating more confidence (Press et al., 1992). As each location on the sky will be measured independently of all other measured by the Planck satellite, each measured value can be tested independently from all other measures - assuming correct calibrations have been applied.

With COBE's differential measurement, no single location in a sky map is measured independently of any others, thus no location should be tested for nonzero value independent of all others. Thus, for COBE, a reliable statistical test should be valid at the level of the sky map rather than individual pixels. Therefore, it is proposed that, the null hypothesis should be extended to image reconstruction by formulating it in terms of sky maps instead of the values of individual pixels (Cover 2006). For COBE the null hypothesis we are interested in rejecting is that there are no anisotropies outside the galactic band.

Underpinning most theories of image reconstruction is the familiar but important concepts of ideal signal and ideal noise. Ideal signal is assumed to maintain exactly the same value over all measurements. Ideal noise is usually assumed to be Gaussian, uncorrelated, be added to the ideal signal (additivity) and maintain exactly the same distribution over all measurements (stationarity). A widely known and very useful property of ideal noise is that it obeys the square-root-of-N rule under averaging, when N is the number of data points. Thus, for ideal noise, the SNR ratio of data can be improved by a factor of 1000 by averaging together 1,000,000 measurements of the same signal. In the official analysis of the COBE data, it appears the temperature variations of the anisotropies from the CMB was assumed to be ideal signal and the instrumentation noise, after calibration corrections, was assumed to be ideal noise.

A widely used measure of the consistency of a model with a data set is the $\chi^2$ value. In the COBE 4 year analysis the models are the possible sky maps and the data are the differential measurements over the 4 years. Because of the nature of the COBE measurements, the differential measurements are often referred to as the time ordered data (TOD). The equation for calculating the $\chi^2$ value is

$$\chi^2 = \sum_t \left[ \frac{D_{ij}(t) - (T_i - T_j)}{\sigma_{ij}(t)} \right]^2 \qquad (1)$$

over the time, t, where $D_{ij}(t)$ is the TOD, $\sigma_{ij}(t)$ is the standard deviation of the noise of the pixel pair, ij, observed at time t (Kogut et al. 1996A). The larger the value of $\chi^2$ of a potential sky map, $T_i$, the smaller the confidence in the fit. Assuming ideal signal and noise, a large value for N, and that the sky map, $T_i$, is the true (but unknown) sky map, the expected value for $\chi^2$ is N, and the standard deviation of $\chi^2$ is $(2N)^{1/2}$ (Press et al. 1992). Thus, for any proposed sky map we can calculate a confidence level for which the sky map is consistent with the data. The confidence level is usually given in terms of the number of standard deviations the $\chi^2$ value of the proposed sky map falls within the expected mean of the $\chi^2$ value. Although, if desired, the confidence level can be expressed as a p-value.

If we could show that all of the sky maps within a specified confidence level of the expected mean of the $\chi^2$ value had anisotropies, then the null hypothesis would be rejected to the specified confidence level. Assuming the confidence level was sufficiently small, the anisotropies would be definitively detected. However the question arises, if we found a single sky map consistent with the data to within the confidence level that had no anisotropies, would the null hypothesis still be rejected?

It is possible to calculate a probability density of a particular sky map directly from the $\chi^2$ using the equation

$$\text{Pr} ob(\chi^2) = k \exp(-\chi^2/2) \qquad (2)$$

(Press et al. 1992 p 820) where k is a constant independent of $\chi^2$. Thus the probability density of any sky map is completely and uniquely determined by its $\chi^2$ measure of consistency with to the data. A consequential property of the relationship between $\chi^2$ and the probability density is that the probability density strictly decreases with increasing $\chi^2$. It is important to note that, although there is sufficient information to assign a probability distribution to the $\chi^2$ value, the TOD does not have sufficient information to assign probabilities to individual sky maps, only probability densities can be assigned. The probability of a model, however, cannot be calculated without prior information (Tarantola 1987). As no reliable prior information exists for the COBE sky maps reliable probabilities cannot be calculated for individual sky maps.

Therefore, if there is at least one sky map, that is consistent with the data with a specified confidence level that does not have anisotropies, it is impossible to assign a probability to the existence of the anisotropies based only on the information in the data alone. Thus, all that can be said is that within the specified confidence level the data is neutral as to the existence of anisotropies. Consequently, any detection of anisotropies, or lack there of, must be at a lower confidence level.

In some circumstances statistical tests designed for the detection of signals in direct measurements, when applied to reconstructed images, will give reliable results. It has been shown that if the forward problem is linear, and has ideal signal and noise, and the reconstruction algorithm is linear with good resolution, such circumstances are achieved (Cover 2006).

## 3. METHOD

The TOD for COBE consists of 3 pairs of channels. The frequency of each pair were 31.5, 53 and 90 GHz. Each channel pair had two polarisations labelled A and B. Channel 31B was not used in this reanalysis because of poor data quality. Bennett et al. (1996A), averaged together the sky maps of the A and B channel pairs after reconstruction to improve the SNR and displayed the results as a sky map for each frequency. In this analysis, the A and B channel sky maps will be presented separately to provide independent confirmation of the results.

For both the official reconstruction and this reanalysis the COBE TOD was fitted to a pixelized sky map, Tj, consisting of 6144 pixels. The fitting was accomplished for each channel using the maximum likelihood criterion.

The form of the COBE TOD data used in the reanalysis were the "pixelized" differential data files from the official 4 year analysis and was downloaded from the Legacy Archive for Microwave Background Data Analysis (LAMBDA) web site maintained by the Goddard Space Flight Center. The pixelized data was calculated from the TOD by averaging together observations that had the same antenna directions in the pixelized sky. The data in the files included corrections for known systematic errors, including corrections to reduce the calibration error from near 3% to 0.7%, and subtraction of a nominal dipole.

The numerical optimization was implemented using the conjugate gradient algorithm (Press et al. 1992). Pixels further than a specified number of degrees from the galactic equator were constrained to zero by a simple method. First, the values for all pixels in all optimisations were set to zero. Second, for all pixels that were to be constrained to zero during an optimization, the corresponding gradient was forced to zero. This ensured that the required pixel values would remain unchanged at zero as desired. To correct for the monopole in each reconstructed sky map, the average value of all pixels more than 30° from the galactic equator was subtracted from every pixel in the respective sky map.

The smoothing and display of the sky maps closely follows the steps described in Bennett et al. (1996B) including smoothing with a 7° Gaussian filter and the use of Mollweide projections. All calculations were performed on a PC with a 1200MHz Athlon CPU with 512MB of memory and written in Java. The time to calculate a single sky map for a single channel was between 5 and 10 minutes.

The official COBE analysis used the normalised $\chi^2$ values in place of the actual $\chi^2$ values. The normalised $\chi^2$ values were calculated by dividing the actual $\chi^2$ values by the number of TOD measurements, N. This convention was used in the reanalysis.

## 4. RESULTS

Figure 1 shows how the normalised $\chi^2$ values varies with an increasing number of constrained pixels for each of the 5 reconstructed channels. The increasing number of constrained pixels from 90° to 24° shows little change in the $\chi^2$ values but below 24° they begin to show a distinct increase.

Table 1 shows comparisons between the sky maps produced by the official 4 year analysis of the COBE data and those reconstructed during the reanalysis from the pixelized data. The recalculated $\chi^2$ values for the official sky maps and the sky maps reconstructed for the 5 channels with all pixels free to vary are within 3 parts per million of each other. The maximum absolute difference and the root-mean-square (RMS) of the difference between the sky maps do not exceed 65.9µK and 24.2µK respectively. The agreement between the $\chi^2$ values of the official and reanalysis sky maps indicates the image reconstruction in the reanalysis is performing in the same way as the official analysis.

Table 2 shows the precise value of the recalculated normalised $\chi^2$ values for the official reconstruction from the official COBE analysis as well as the values from the reanalysis at 90°, 24° and 0°. The percentage increase in the $\chi^2$ values from the 90° to the 24° sky maps ranged between 0.21% and 0.24%.

Also in Table 2, the $\chi^2$ values at 0° indicate how consistent the data from each channel is with a sky map with no signal. In both Fig. 1 and Table 2, all channels show a clear increase in the $\chi^2$ values at 0° as compared to 90°. Channel 31A shows the highest increase of 3.3%, suggesting the poorest fit to no signal and thus the highest signal from the galactic band. Channels 53A and 53B show intermediate increases and channels 90A and 90B show the lowest at 0.61% and 0.74% respectively.

## 5. DISCUSSION

From the point of view of image reconstruction, the COBE sky map reconstruction problem is a particularly interesting one for several reasons. First, the signal due to the reported anisotropies in the TOD is only 1 part in a thousand of the instrumentation noise. Second, with the galactic band radiating a much larger signal than the reported anisotropies, the COBE image reconstruction is particularly susceptible to reconstruction artefacts. Third, the maximum likelihood reconstruction algorithm used in the official COBE analysis does not take into account the variety of other sky maps that are also consistent with the same TOD. In addition, there are many papers that describe the analysis of the COBE data in detail, thus facilitating any reanalysis.

As an added bonus, a few hours spent acquiring repetitive circular scans with the WMAP satellite, which is currently measuring the CMB, would provide definitive detection of any 10ppm anisotropies without the use of image reconstruction (discussed below). Also, before the end of 2008 the Planck satellite will be launched providing a second independent measure of any CMB anisotropies.

**5.1 Averaging instrumentation noise**

While this paper reanalyses the COBE data from the point of view of image reconstruction, the importance of the ideal noise assumption to the official COBE data analysis can be demonstrated in a simpler way. The instrumentation flight noise of the 5 channels reanalysed in this paper ranged from 23.13mK to 58.27mK for each 0.5s observation in the TOD (Bennett et al, 1996A). The amplitude of the reported anisotropies was roughly 30µK. Therefore, to reliably detect the anisotropies, the instrumentation noise must be reduced by at least a factor of 1000.

The official COBE analysis accomplished all its noise reduction by averaging. Explicit averaging was performed when the pixelized data was calculated from the TOD. Implicit averaging was performed as part of the sky map reconstruction from the pixelized data since the maximum likelihood reconstruction can be represented as a matrix multiplication. To reduce the instrumentation noise by a factor of 1000 would require the explicit averaging of 1,000,000 points assuming ideal noise. However, the implicit averaging taking place during the image reconstruction both adds and subtracts the weighted averages of the pixelized data. For a complete analysis, the official COBE analysis needed to show that the instrumentation noise behaved as ideal noise for many times more than 1,000,000 averages.

The 3.0% uncertainty in the calibration of TOD may have caused the combination of the signal from the galactic band and the instrumentation noise to deviate from the square-root-of-N rule, which is characteristic of ideal noise, at a values less than 1,000,000 averages. The official COBE analysis only provided evidence the square-root-of-N rule held to about 6,000 averages (Kolgut et al. 1996A, Fig 1). As the artefacts due to the calibration error would yield defuse signals spread across the whole sky, official COBE analysis may have detected artefacts due to non ideal noise rather than anisotropies in the CMB.

**5.2 Rejecting the null hypothesis**

To be able to reject the null hypothesis that there are no anisotropies outside the galactic band, we need a good estimate of the standard deviation of the $\chi^2$ values over the 4 years of data acquisition. The noise of the TOD is reported in the official COBE analysis to be uncorrelated Gaussian (Kogut et al. 1996A), and the official analysis seems to have assumed it to be ideal. For example, there were 218,586,649 measurements for channel 53B. According to the probability theory presented in the Theory section of this paper, and the assumption of ideal noise, the standard deviation of the normalised $\chi^2$ values would be roughly 0.000096.

The normalised $\chi^2$ values in the official analysis actually varied over the channels from 0.93 to 1.18 as they do in this reanalysis. This discrepancy is discussed in Bennett et al. (1996B) and is attributed to a change in the noise of the channels from pre-flight measurement to in-flight measurements. However, it is the change in the standard deviation of the noise over the 4 year mission, not the change at launch, that is important to the analysis.

From Table 2, the differences between the $\chi^2$ values of the official sky maps and the sky maps when all pixels outside of 24° of the galactic equator were forced to zero ranged between 0.20% and 0.27%. If the ideal noise standard deviation for $\chi^2$ is used, the null hypothesis is rejected by $28\sigma$ (0.0027/0.000096) for channel 53B. This very high confidence level of detection is due to all the pixels outside of 24° of the galactic equator being included in the statistical test. However, as mentioned above, the ideal noise assumption has only been confirmed to about 6,000 averages in the official analysis (Kolgut et al. 1996A, Fig 1). Therefore, the null hypothesis, that there are no anisotropies outside the galactic band, can not be rejected at any significant level for any of the channels.

**5.3 Spectral signature**

A key argument presented in the official COBE analysis was that the signal outside the galactic band was indeed due to anisotropies because the spectral signature of the anisotropies corresponded to that of a black body with the temperature of the CMB (Smoot et al 1992). For the CMB the signal intensity is several times larger in the 53GHz band than the 31.5GHz band and several times larger again in the 90GHz band. However, an early step in the official analysis of the TOD was to convert the data from units of signal intensity to temperature assuming an ideal black body spectrum. Thus, the spectral signature of black body radiation for COBE's three frequency bands was that they all have the same temperature. As a reconstruction artefact that can generate a signal of roughly the same intensity in each of the three frequency bands is more likely, the conversion from units of signal intensity to temperature may have made the CMB spectral signature more easily mimicked by artefacts.

The spectral signature was also used in the official analysis to rule out the possibility that the reported anisotropies were due to signal from the galactic band misplaced by image reconstruction. As is well known, the galactic band has a spectral signature that is distinctly different from the CMB thus allowing minimally processed multiband signals to easily distinguish between signals from the galactic band and the CMB. However, the extensive processing of the TOD in the official analysis introduced the possibility that the spectral signature of the galactic band may have been distorted to the point were it might have mimicked a black body spectral signature. This possibility was not considered in the official analysis.

As mentioned above, the TOD gain drift at measurement (also called the calibration error) was near 3% and was estimated in the official analysis to have been reduced to 0.7% by data processing techniques. For example, pixel 4873 of channel 53B is the brightest pixel of the official COBE sky maps and has a temperature of 1.82mK. Thus the 3.0% drift in the calibration at measurement introduces a drift of up to 55$\mu$K in the TOD, roughly double the size of the anisotropies reported by COBE. Therefore artefacts due to the gain drift have roughly the correct size to have been mistaken for anisotropies in the official analysis.

The reconstruction algorithm used in the official COBE analysis is often treated as a simple maximum likelihood reconstruction. A useful property of a simple maximum likelihood reconstruction is that it can be represented as a matrix multiplication making its properties much simpler to characterise. This simplification was used earlier in this paper to demonstrate the COBE TOD was also consistent with no anisotropies. However, the official COBE analysis also incorporated the calculation of the gain drift corrections into the maximum likelihood reconstruction thus making the algorithm nonlinear (Kogut et 1996A). The nonlinear nature of the official reconstruction introduces the possibility that the spectral signature of misplaced signal from the galactic band was modified by the combined image reconstruction and gain drift correction.

The possibility that the official COBE analysis altered spectral signatures from the galactic band to that of the CMB is of particular concern because, as mentioned above, the amplitude of the artefacts due to the gain drift would be roughly the same size as the anisotropies reported by COBE. In addition, the official analysis did not considered the possibility that combining the image reconstruction with the gain drift calculation may alter the spectral signatures.

The WMAP satellite had similar gain drift errors to that of COBE and corrected for them in a similar manner (Hinshaw et al. 2003). Therefore it is possible WMAP would produce similar artefacts mimicking anisotropies in the CMB. All other reported detections of anisotropies in the CMB used balloon or ground based measurements that had to detect the anisotropies through the earth's atmosphere. The earth's atmosphere introduces additional calibration errors that are more difficult to correct for than those of spaced based measurements. Avoiding the additional calibration errors is the primary reason so much time effort and money has been spent on spaced based measurements of the reported anisotropies in the CMB. Thus, the balloon and ground based measurements are even more susceptible to the artefacts due to gain drift that might mimic anisotropies.

**5.4 Repetitive circular scans**

Repetitive circular scans are different than the intersecting circular scans of the COBE and WMAP satellites. The intersecting circular scan pattern is achieved by rotating each of the satellites about an axis while slowly precessing the axis of rotation. The WMAP satellite scans a full circle in slightly more than 2 minutes. If the precession of the WMAP satellite was stopped and the axis of rotation was set parallel to the earth-sun axis, WMAP's two beams would follow each other's track through the sky thus following a repetitive circular scanning pattern. Analysis of measurements made during repetitive circular scan should be much easier if each measurement lines up exactly from rotation to rotation.

Repetitive circular scans have substantial advantages when dealing with the problem of gain drift. If the drift is slow when compared to the rotation rate of the satellite, then it can be assumed to be constant during a single rotation. This assumption makes any image analysis much more predictable because the first step is usually averaging the measurements from many rotations into one. If the gain can be reliably assumed to be constant during each rotation, even if there is a substantial change in the gain from one rotation to the next, the averaged circular scan of the sky will have the same, but unknown, gain error for all measurements. Thus the shape of any features in a circular scan will be unchanged by the gain drift.

Repetitive circular scans can be used by both WMAP and Planck to detect anisotropies. The signal to noise ratio of both satellites is sufficient to allow the detection of 10ppm anisotropies of the angular size reported by COBE with only a few averages. The single beam of Planck makes image reconstruction unnecessary to detect the anisotropies. As WMAP makes differential measurements, the simplest way to detect anisotropies is to only use observations when both beams are outside the galactic band. The difference between the two beams will then be within root two of the approximate amplitude of the anisotropies.

Averaging hundreds or thousands of repetitive circular scan rotations with either satellite should allow anisotropies much smaller than 10ppm to be detected if necessary. Limited statistical properties of any anisotropies detected might also be determined. The high signal to noise ratio achievable with repetitive circular scans may also aid the study of other properties of the CMB such as polarisation.

**5.5 Independent verification of anisotropies**

The high correlation between the results of the various groups measuring the anisotropies in the CMB means that there must be information shared among the analyses. But this information does not have to be the signal from anisotropies as has been assumed. Standard practices in statistical analysis required that, as part of any claims of a detection by correlation, any other confounding sources of the correlation need to be discussed and ruled out (Altman 1991). The literature on the detection of the CMB anisotropies has given little attention to the possibility of confounding correlations.

One possible source of shared information between COBE and other analysis is the use of results from the COBE official analysis in the analysis of other groups. For example, if the COBE sky maps were used to initialise the nonlinear reconstruction algorithms used by other groups such as WMAP, this may introduce confounding correlations between the data sets. Another example would be the use of the COBE dipole in the analysis of data sets from other groups. Other similarities in the design of the TOD processing pipelines may also introduce confounding correlations.

**6 CONCLUSIONS**

The statistical concept of rejecting the null hypothesis was extended to image reconstruction by asking the question of whether the COBE TOD was also consistent with no anisotropies outside the galactic band. Provided the data was assumed to obey the ideal noise assumption to more than 200,000,000 averages, as it appears was done in the official COBE analysis, the null hypothesis can be rejected at 28σ providing a definitive detection. However, the official analysis only provided evidence of the ideal noise assumption holding to 6,000 averages. Thus the null hypothesis cannot be rejected as the TOD is also consistent with no anisotropies.

The 3.0% calibration error in the TOD, perhaps compounded by signal from the galactic band, may well have caused the TOD to substantially deviate from the ideal noise assumption. In addition, the combining of the image reconstruction with the gain drift correction in the official COBE analysis resulted in a nonlinear reconstruction algorithm that may have altered the spectral signature of the galactic band so it appears similar to that of the CMB. As the reconstruction algorithm used to generate the sky maps assumes ideal noise, it appears that it may have reconstructed non ideal noise caused by the calibration uncertainty in a way that could be misinterpreted as anisotropies in the CMB.

Nothing in this paper should be taken as a definitive statement that anisotropies in the CMB at 10ppm or less do not exist. Rather, this paper questions the completeness of the official COBE analysis in ruling out possible sources of signal that could be mistaken for anisotropies.

A few hours of repetitive circular scans with the WMAP satellite or direct measurements with the single beam Planck satellite should provide definitive evidence as to the existence or non existence of the 10ppm anisotropies without the complications introduced by image reconstruction.

**References**

Altman DG. 1991, Practical statistics for medical research. Chapman and Hall, London, England.

Barnes C., et al., 2003, "First year Wilkinson microwave anisotropy probe (WMAP) observations: Galactic signal contamination from sidelobe pickup," ApJS, 148, 51

Bennett C.L., et al. 1996A, "4-Year COBE DMR Cosmic Microwave Background Observations: Maps and Basic Results," ApJ, 464, L1

ed. Bennett C.L., Leisawitz D, Jackson P.D., 1996B, "COBE-DMR Four-year project data sets (PDS), analyzed science data sets (ASDS), and galactic coordinate data sets explanatory supplements" COBE Ref Pub. No. 96-B (Greenbelt, MD; NASA/GSFC), available in electronic form from the NSSDC.

Bennett C.L., et al. 2003, "First year Wilkinson microwave anisotropy probe (WMAP) observations: preliminary maps and basic results," ApJS, 148, 1

Cover K.S., 2006, "A multiexponential reconstruction algorithm immune to false positive peak detection", Rev. Sci. Instrum., 075101.

Hinshaw G., Barnes C., Bennett C. L., Greason M. R., Halpern M. , et al. 2003. "First YearWilkinson Microwave Anisotropy Probe (WMAP) Observations: Data Processing Methods and Systematic Errors Limits," ApJ, 148, 63-95.

Kolgut A., Smoot G.F., Bennett C.L., Wright E.L, et al., 1992. "COBE differential microwave radiometers: preliminary systematic error analysis", ApJ, 401, 1-18

Kogut A., Banday, A. J, Bennett C. L., Gorski, K. M., Hinshaw G., Jackson P. D., Keegstra, P., Lineweaver C., Smoot, G. F., Tenorio, L., Wright E. L., 1996A, "Calibration and systematic error analysis for the COBE DMR 4 year sky maps," ApJ, 470, 653-673

Kogut A., Banday A.J., Bennett C. L., Gorski K. M., Hinshaw G., Smoot G. F., Wright E. L., 1996B, "Microwave emissions at high Galactic latitudes in the four-year DMR sky maps," ApJ, 464, L5-L9

Press W.H., Teukolsky S.A., Vetterling W.T., Flannery B.P., 1992, Numerical recipes in C. Cambridge Univ. Press, Cambridge

Smoot G. et al., 1992, "Structure in the COBE differential microwave radiometer: $1^{st}$ year maps," ApJ, 396, L1

Spergel D.N. et al., 2003, "First year Wilkinson microwave anisotropy probe (WMAP) observations: determination of cosmological parameters," ApJS, 148, 175

Starkman G. D. and Schwarz D. J., 2005, "Is the universe out of tune?", Scientific American, 293, 48

Tauber J. A., 2004, "The Planck mission", Advances in space research, 34, 491

Tarantola A., 1987, Inverse Problem Theory, Elsevier, Amsterdam

**Figure Captions**

**Figure 1.** Graph of the $\chi^2$ values versus the number of degrees from the galactic equator beyond which the pixels in the sky maps were constrained to zero. As can be seen from the plots, there was virtually no change in the $\chi^2$ values between 90° and 24°. The horizontal axis is labelled in units of degrees but is spaced by the percentage of pixels that are not constrained to zero.

**Figure 2.** Sky maps reconstructed as part of the reanalysis with none of the 6144 pixels constrained to zero and thus matching the official COBE reconstructions. The sky maps clearly show that the signal from the galactic band is many times larger than that of the apparent anisotropies.

**Figure 3.** Sky maps reconstructed as part of the reanalysis with pixels further than 24° from the galactic equator constrained to zero. These sky maps are also consistent with the COBE TOD.

**Tables**

**Table 1.** Comparison of the sky maps from the official 4 year analysis (Bennett et al 1996) and the sky maps reconstructed during the reanalysis. The maximum absolute difference and root-mean-square (RMS) difference between the sky maps are over all of the 6144 pixels.

| Channel | Number of observations | Official Normalized $\chi^2$ (Bennett et al. 1996) | Reanalysis Normalized $\chi^2 (b<90°)$ | Sky Maps Maximum Absolute Difference (mK) | Sky Maps RMS of Difference (mK) |
|---|---|---|---|---|---|
| 31A | 168,132,225 | 1.12118354 | 1.12118002 | 0.0659 | 0.0242 |
| 53A | 218,557,665 | 1.05410507 | 1.05410233 | 0.0337 | 0.0131 |
| 53B | 218,586,649 | 1.18123722 | 1.18123414 | 0.0337 | 0.0131 |
| 90A | 218,398,620 | 0.93740230 | 0.93740007 | 0.0324 | 0.0078 |
| 90B | 218,360,156 | 1.17461294 | 1.17460992 | 0.0260 | 0.0070 |

**Table 2.** The $\chi^2$ values of the sky maps reconstructed during the reanalysis for no pixels constrained to zero (b<90°), pixels outside 24° of the galactic plane constrained to zero (b<24°), and all pixels in the sky maps constrained to zero (b=0°). (The second column is reproduced from Table 1).

| Channel | Reanalysis Normalized $\chi^2$ (b<90°) | Reanalysis Normalized $\chi^2$ (b<24°) | Reanalysis Normalized $\chi^2$ (b=0°) | Differences in Normalized $\chi^2$ (b=24° minus b=90°) | Differences in Normalized $\chi^2$ (b=0° minus b=90°) |
|---|---|---|---|---|---|
| 31A | 1.12118002 | 1.12359839 | 1.15818237 | 0.00241837 | 0.03700235 |
| 53A | 1.05410233 | 1.05668081 | 1.07795510 | 0.00257848 | 0.02385277 |
| 53B | 1.18123414 | 1.18397989 | 1.20146171 | 0.00274575 | 0.02022757 |
| 90A | 0.93740007 | 0.93941973 | 0.94313518 | 0.00201966 | 0.00573511 |
| 90B | 1.17460992 | 1.17721477 | 1.18330889 | 0.00260485 | 0.00869897 |

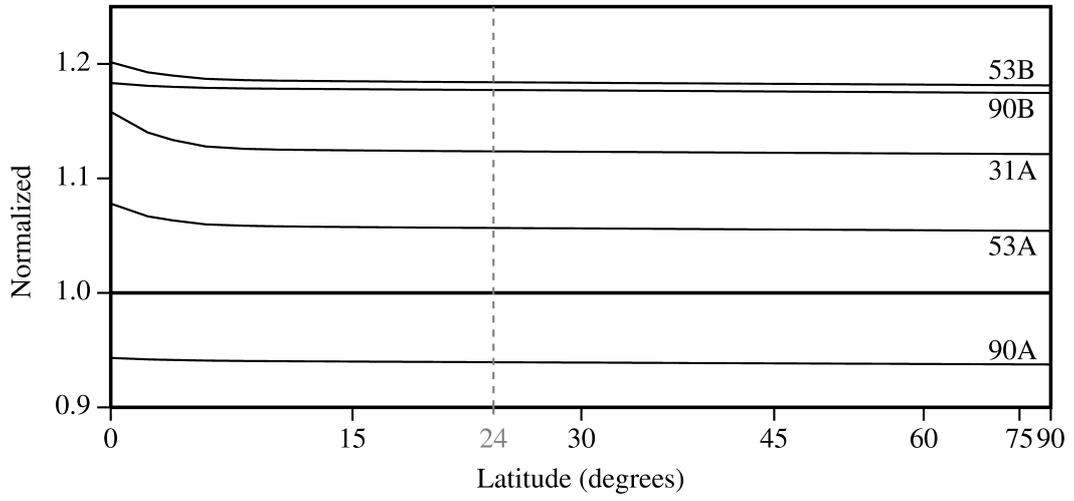

Figure 1

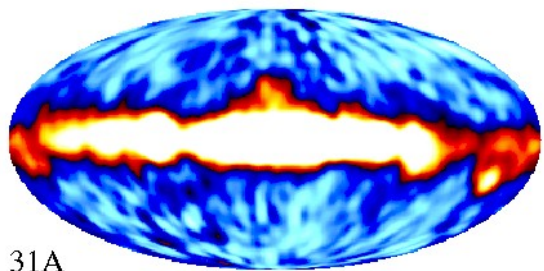
31A

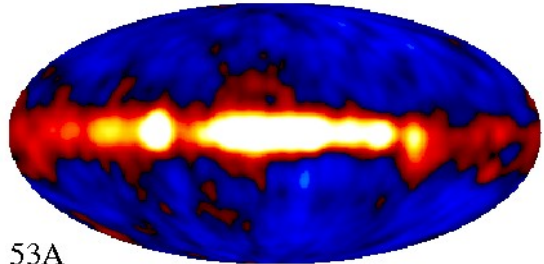
53A

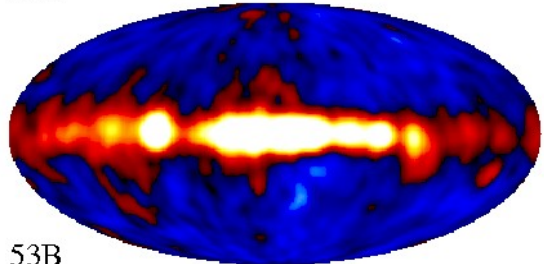
53B

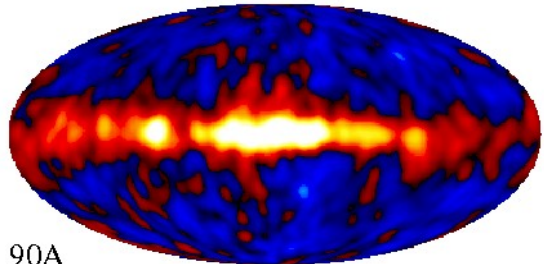
90A

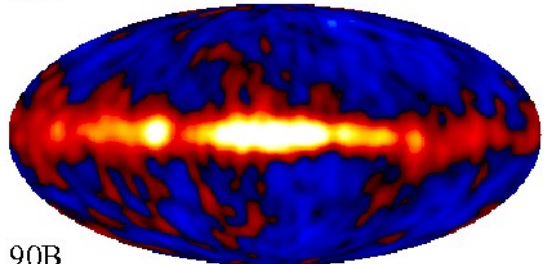
90B

Figure 2

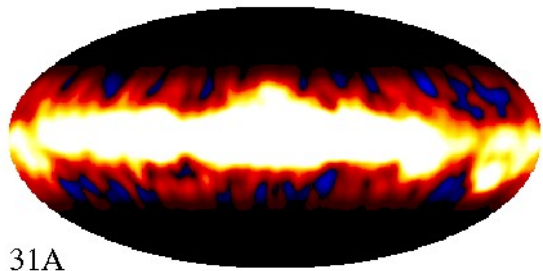

31A

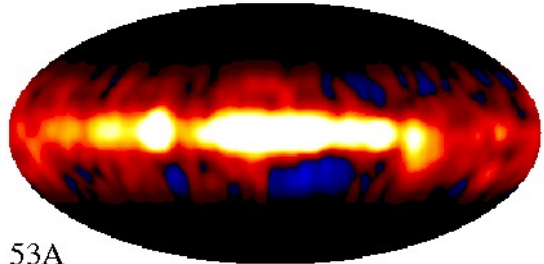

53A

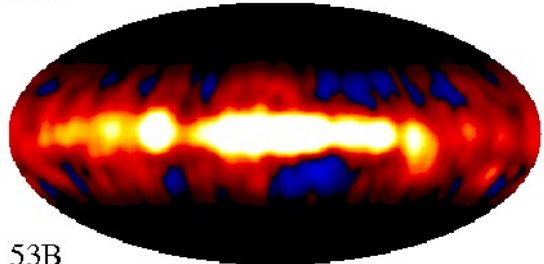

53B

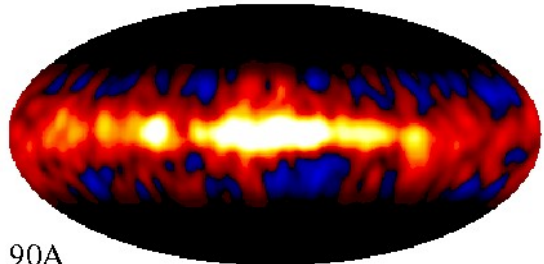

90A

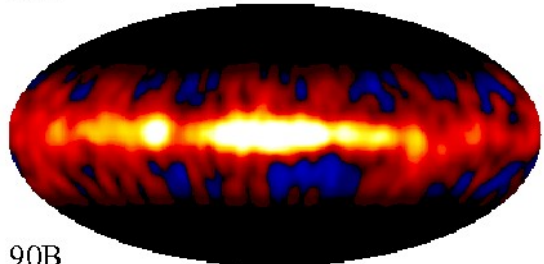

90B

Figure 3